| Abstract | For the past twelve years, cold plasmas (i.e., weakly ionized gas) have been positioned as a breakthrough technology for treating cancer thanks to their antitumor properties. The innovation of *ad hoc* plasma sources and personalized protocols appears crucial to treat cancers with a very poor prognosis. This is the case for cholangiocarcinoma (CCA), a biliary tract cancer, whose treatment with cold plasma is envisioned but requires the innovation of catheters and endoscopic devices for local therapies. Before conducting clinical trials, the performances and limitations of cold plasma endoscopy must be evaluated in terms of safety for both the patient and clinician as well as in terms of therapeutic efficacy. These objectives are pursued in the present work, in which a transferred plasma catheter is used, powered by 8-10 kV, 1-2 µs wide, 5-10 kHz repeated pulses, allowing guided streamers to be transferred over lengths of at least 2 m. In a first step, the catheter is utilized without the duodenoscope and directly inserted into an artificial model reproducing the topography and the electrical response of the biliary tree. This model allows to validate the technical feasibility of the technology and to demonstrate the absence of electrical and thermal risks. Indeed, the voltage and current deposited are as low as 3.98 V and 1.19 mA (RMS values) respectively, while the temperature locally increases from 23°C to 27 °C. In a second step, the catheter is inserted into the duodenoscope, the whole being applied to a porcine anatomical model. After passing through the esophagus, stomach and duodenum, the distal part of the duodenoscope is placed at the entrance to the papilla so that the catheter can easily enter the choledoc and then the common bile duct. Interestingly, the electrical power values deposited are of the order of 100 mW especially because the current values are at least 10 times higher. These more elevated values of electrical parameters but also electromagnetic effects are discussed considering physical aspects like eddy currents. The absence of electrical and thermal risks is demonstrated and consolidated by the IEC standards for medical devices. In a third time, we demonstrate that the cold plasma catheter can induce antitumor effects on *in vitro* experimental models of human CCA. The methodology proposed in this article validates the relevance of cold plasma endoscopy as a potential local treatment for cholangiocarcinoma and allows bridging cognitive and patient-oriented research. |
|---|---|

# I. State of art

## I.1. Cholangiocarcinoma: a rare cancer with poor prognosis

### I.1.1. Cancer prognosis

Cancer is one of the leading causes of death in all countries. In 2020, there were 19.3 million new cases and 10.0 million cancer-related deaths worldwide [1]. By 2040, the number of new cancer cases per year is expected to rise to 29.5 million and the number of cancer-related deaths to 16.4 million [2]. In general, the countries with the highest cancer rates are those where populations benefit from a high life expectancy, a high level of education and a high standard of living. However, there are some exceptions such as cervical cancer whose incidence rate is higher in low-GDP countries [3]. Besides, within the same country, the survival rate of patients can highly change depending on the type of cancer. Hence, in France, the standardized net survival (SNS) which considers the effect of age by standardization at 5 years varies from 96% for thyroid cancers to 10% for mesothelioma of the pleura. Cancers with an unfavorable prognosis represent, in terms of incidence, 32% of solid tumors in men and 19% in women, as well as 7% of hematological malignancies in men and 9% in women [4], [5]. Conversely, cancers with a favorable prognosis represent 40% of solid tumors in men and 55% in women, as well as 45% of hematological malignancies in both men and women [6].

Among cancers of poor prognosis, the rare and aggressive ones present often tough challenges that make them more difficult to treat than common tumor types. For a rare cancer case, doctors may not have a standard FDA-approved (and/or EMA-approved) therapy to help guide them in clinical decision-making. In aggressive cancer cases, cancer cells can often become resistant to standard treatment options, and patients may therefore exhaust these options very quickly. For these reasons, finding new treatments is an absolute emergency for patients with rare and/or aggressive cancers such as cholangiocarcinoma (CCA): the cancer of the biliary ducts.

### I.1.2. CCA : a biliary tract cancer

As sketched in Figure 1, the biliary tree is a series of gastrointestinal ducts allowing modification and transport of bile primarily synthesized by hepatocytes in the liver to be concentrated and stored in the gallbladder, prior to release into the duodenum at mealtime. The bile canaliculi gives rise to the canals of Herring, at the origin of the bile ducts, then to the intrahepatic bile ducts from which the right and left hepatic ducts join outside the liver to form the common bile duct, then the cystic duct and the choledoc which empties into the duodenum through the great caruncle via the ampulla of Vater and the sphincter of Oddi [7].

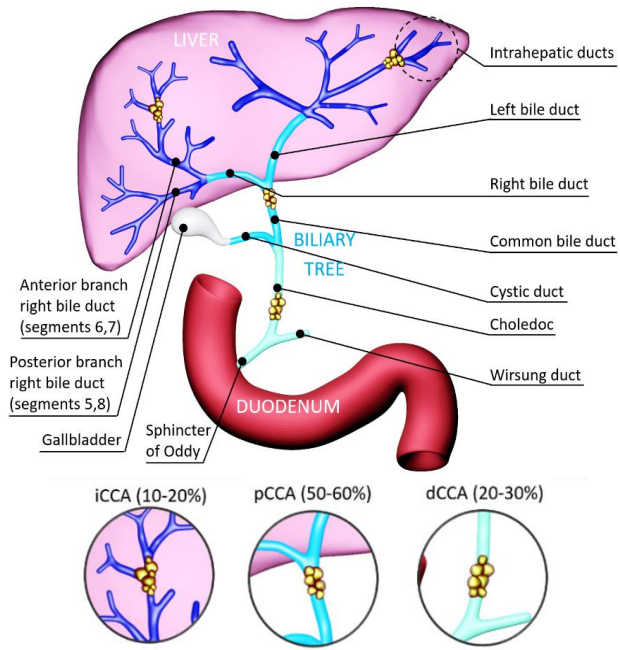

*Figure 1. Anatomical representation of the biliary tree, its environment (liver and duodenum) and the cholangiocarcinoma sites (intrahepatic, perihilar and distal).*

CCA corresponds to a cluster of highly heterogeneous malignant tumors that can develop at every point of the biliary tree, regardless to their size or location: in the liver (intrahepatic, iCCA), at the hilum (perihilar, pCCA) or distally (distal, dCCA). Their occurrences are 10%, 50% and 40% for iCCA, pCCA and dCCA respectively. According to the new nomenclature, these last 2 sites are considered as extrahepatic CCAs (eCCA). The mutational profile of CCAs also varies:

(i) iCCA have genetic alterations in IDH-1 (15-20%) and FGFR-2 (fusion) (10-15%) that can be targeted by inhibitors, currently approved by the FDA and/or the EMA. Phase II and III clinical trials are currently being conducted with pemigatinib and ingratinib (FGFR-2 inh), and ivosidenib (IDH-1 inh) [8], [9]. Pemigatinib is now approved for advanced CCA patients with *FGFR2* rearrangements previously treated with GEMCIS.

(ii) eCCA also have genetic alterations, some of which are known to be targeted by inhibitors already present on the drug market [10].

CCA is the second type of primary liver cancer (~15% of all cases) and currently accounts for about 3% of gastrointestinal carcinomas [11]. To date, this rare and aggressive cancer represents nearly 2% of annual global cancer mortality. The main risk factors of this malignancy are chronic inflammatory diseases of the bile ducts, parasitic infections (in Asia), alcohol consumption, and infections with hepatitis B and C viruses. However, in 50% of cases, there is no risk factor identified. One of the main histological features of CCA is its extremely fibrous stroma creating a physical barrier and thereby decreasing the ability of chemotherapies or targeted therapies to reach tumor cells [11]. CCA malignancies are asymptomatic and only suspected due to the occurrence of jaundice or abnormalities on liver function tests, consequences of defective bile flow caused by tumor obstruction in the bile ducts. In consequence, the diagnosis comes at an advanced stage, locally advanced or metastatic, hence contributing to the dark prognosis of CCA, with a 5-year overall survival rate less than 30% [12]. CCA continues to have a poor prognosis, with little improvement over the past 30 years in terms of resectability or early diagnosis, with relative survival rates at 1, 3 and 5 years of 25%, 10% and 7%.

### I.1.3. CCA conventional treatments

Today, CCA is treated following different medical approaches and depending on tumor progress stages, not all of them are relevant. While some treatments are well established to provide curative care (resection, adjuvant chemotherapies) and palliative care (chemotherapies), a panel of optional treatments can also be proposed, as reported in Table 1.

Endoscopically applied plasma therapy has multiple advantages over the approaches presented in Table 1. First, it is minimally-invasive and can therefore address a greater number of patients, unlike surgery and liver transplantation. While biliary drainage does not induce an antitumor effect, cold plasma endoscopic therapy can induce tumor cell apoptosis without inducing harmful systemic effects on the rest of the human body as is the case with chemotherapy or targeted drug therapy. Moreover, its therapeutic action is not only based on the exploitation of a single physical property as proposed by PDT (radiation property) or endobiliary thermal ablation (temperature property). Cold plasma can effectively simultaneously induce pulsed electric fields for cell permeation and produce radicals/Reactive Oxygen Species (ROS) for oxidative stress enhancement but also generate UV radiation for bacteria/virus decontamination. Furthermore, cold plasma endoscopic therapy can activate immune cells from the tumor microenvironment just as conventional immunotherapy would. Finally, the combination of all these properties within cold atmospheric plasma can drive therapeutic synergies that none of the conventional approaches can implement alone.

|  | Principle | Limitations |
|---|---|---|
| **Surgery** | • For very small bile duct cancers, surgery involves removing part of the bile duct and joining the cut ends.<br>• For more-advanced bile duct cancers, nearby liver tissue, pancreas tissue or lymph nodes may be removed as well. | Surgical intervention is the only curative treatment but only possible on 20-30% of patients. Relapse rates are high and range from 40-70% [13]. |
| **Chemo-therapy** | • Chemotherapy is an option for people with advanced CCA to help slow the disease and relieve signs and symptoms. Chemotherapy drugs can be infused into a vein so that they travel throughout the body. The drugs may also be administered in a way so that they are delivered directly to the cancer cells (intra-arterial chemoembolization).<br>• First-line treatment (GEMCIS) and second-line treatment (FOLFOX) [13] | • This approach remains palliative<br>• This systemic approach is not a cure and only allows a temporary control of the disease: survival rate is increased by only a few months, with non-negligible side-effects. |
| **Liver trans-plantation** | Surgery to remove the liver and replace it with one from a donor may be an option for people with iCCA (combined to neoadjuvant chemotherapy) [14] and pCCA [15]. | • Curative treatment with restrictive patient selection.<br>• Risk of recurrence. |
| **Immuno-therapy** | Immunotherapy may be an option for CCA of advanced cancer if other treatments have failed. Treatments targeting immune checkpoint inhibitors such as PD-1, PD-L1, CTLA-4 or a combination of them are currently investigated in clinical trials. [16] | Body's disease-fighting immune system may not attack cancer because cancer cells produce proteins that help them to escape from the immune system. As part of clinical trials, monotherapies with immune checkpoint inhibitors show poor efficacy to date [16]. |
| **Targeted drug therapy** | Focus on molecular abnormalities present within cancer cells. By inhibiting these abnormalities, targeted drug treatments cause cancer cells to die. | Therapy that is not systematically effective against CCA as it depends on genetic alterations (personalized medicine). |
| **Radiation therapy** | Therapy which relies on high-powered energy beams from sources such as X-rays and protons to kill cancer cells. Radiation beams can be focused onto the body (external beam radiation). It may also involve placing radioactive material inside the body near the site of CCA (brachytherapy). | Survival rate is increased of only a few months [17] |
| **Endobiliary thermal ablation** | Radiofrequency currents locally confined are used to thermally ablate tumor tissues of the bile duct without damaging its structural walls thereof. This procedure can be achieved either endoscopically or percutaneously. It contributes to biliary draining by setting stents (biliary prostheses) that can lift the biliary stenosis [18]. Medical devices marketed by Boston Scientific and Taewong companies | Although commercially available, radiofrequency has not been conclusively proven to be of current use in the therapeutic arsenal of CCA, as the human studies are retrospective, nonrandomized and of small scale. |
| **Biliary drainage** | Biliary drainage is a procedure to restore the flow of bile. This may involve placing a thin tube in the common bile duct to drain bile or performing bypass surgery to redirect bile around the cancer and stents to hold open a collapsed bile duct caused by cancer. | Biliary drainage is used as a preoperative treatment [19]. Signs and symptoms of CCA are relieved but remain uncured. |
| **Photodynamic therapy (PDT)** | A light-sensitive chemical is injected into a vein and accumulates in the fast-growing cancer cells. Laser light directed at the cancer causes a chemical reaction in the cancer cells, killing them. | Multiple treatments are required. Today, PDT is no longer performed [20]. |

*Table 1. Therapeutic options for treating cholangiocarcinoma, either routinely or as part of clinical trials at hospital.*

## I.2. Plasma oncology: a disruptive technological approach

### I.2.1. What is cold plasma ?

Cold plasma are weakly ionized gases composed of active species (electrons, ions, metastables, radicals) that present energetic, radiative, chemical, gas dynamics and electric field properties. Today, their interaction with biological systems (cells, tissues, tumors) is studied to address medical issues such as blood clotting, wound healing, dentistry, repair surgery, cosmetics, infectious / inflammatory diseases and oncology [21], [22].

### I.2.2. Antitumor effects induced by cold plasma

*In vitro* studies have brought to light several biological mechanisms induced by cold plasma. One of the most important is to trigger apoptosis in cancer cells, primarily by supplying radicals and reactive oxygen species, with the advantage to deliver them several millimeters into tissues [23]. Hence, in the case of malignant melanoma cells, the plasma-generated ROS can release TRX1 from ASK1 which then phosphorylates p38 and drives to apoptosis after activation of signaling cascade [24]. These *in vitro* studies are also supported by *in vivo* campaigns (preclinical models) although such works represent roughly 5% of the articles published in the field of cold plasma/oncology so far [25]. Recently, anti-neoplastic effects of plasma have been demonstrated on murine subcutaneous tumor models of melanoma [26], breast [27], prostate [28] and bladder [29]. Interestingly, cold plasma appears as a selective treatment, i.e. able to reduce the viability of malignant epithelial cells without altering healthy ones. Hence, in the case of prostate cancer, non-tumor cells P69 are less sensitive to cold plasma (cell viability = 67%) compared to tumor cell lines LNCaP and PC3 (cell viability of 29% and 23% respectively) [28]. Overall, this selectivity seems determined by the cells or tissues properties (e.g. membrane composition, antioxidant system) rather than by plasma treatment [30].

### I.2.3. Cold plasma applied to CCA

So far, very little work has been devoted to the treatment of CCA by cold plasma. The LPP and the CRSA laboratories (Paris, France) have worked on this tumor model since 2018. First, the two teams have utilized a cold plasma jet (Plasma Tesla Jet, PTJ [31]) to treat ectopic models of CCA subcutaneously inoculated into mice [32]. In addition to inducing significant antitumor effects, this plasma source has not induced toxic effects of skin tissues (no burnings, no tissue necrosis and no alteration of epithelial cells or collagen structure). Hence, two local applications of cold plasma (1 min each) on the tumor reduces by 30% its volume compared to non-treated mice. Consistently, tumors from the control group exhibit larger size and vascularization features while tumors from the plasma group appear smaller and undamaged [32]. Analysis of tumor microenvironment (from mice origin) shows a higher mRNA expression of Agre1 (coding for F4/80 macrophage protein) in plasma group, indicating an enhanced recruitment and/or proliferation of tumor-associated macrophages (TAM). Besides, cold plasma enhances the mRNA expression of two major regulators of monocyte chemotaxis and macrophage trafficking: CCL2 and CCR2 (receptor of CCL2) respectively [31]. CCL2 and CCR2 rates exhibit an increased recruitment of TAM in the plasma group compared with the control group [31]. These results suggest that exposure to cold plasma stimulates the differentiation of monocytes into macrophages. More recently, O. Jones *et al.* demonstrated the interest of combining cold plasma with conventional FOLFIRINOX : a systemic chemotherapeutics used to treat CCA. They show that FOLFIRINOX can be administered at lower doses to reduce its toxicity burden while cold plasma is combined to obtain the same therapeutic efficiency [33].

Due to its location in the bile ducts, cold plasma treatment of CCA can only be envisioned if an endoscopic approach is feasible. With a view to transferring technology from the bench to the patient's bedside, the LPP and CRSA teams have adapted and miniaturized the PTJ device to adapt it to the operating channel of a conventional duodenoscope. This therapeutic approach is primarily aimed at treating eCCA for at least two reasons: (i) its prevalence is 85-90% compared to only 10-15% for iCCA, (ii) eCCA is implanted in the bile ducts of larger diameter, namely hilar bile ducts, cystic duct, common bile duct and common hepatic duct, which facilitates the delivery of cold plasma on tumor sites. Besides, it is worth notifying that cold plasma endoscopic therapy offers several advantages over targeted systemic therapies: (i) it is minimally invasive, (ii) it allows direct locoregional treatments under imaging, regardless the mutational status of the tumor, hence minimizing secondary systemic effects, (iii) it can be combined with systemic therapy whether chemotherapy or targeted therapy as part of a neoadjuvant or adjuvant treatment.

### I.2.4. From cold plasma jets to cold plasma catheters

Cold plasma can be easily generated in ambient air using atmospheric pressure plasma jets (APPJ): devices supplied with a gas flow (typically a noble gas with/without reactive gas) to ensure the propagation of the ionized gas several centimeters away from the interelectrode space. Most AAPJs have two electrodes whose electrical potentials are known and fixed: an electrode polarized at high voltage and a counter-electrode that is grounded. As an example, Figure 2a shows an APPJ where electrical energy is delivered through two outer ring-electrodes so that plasma is never in direct contact with them. This type of APPJ has been successfully used for pre-clinical applications [32], [31]. In some AAPJ devices, cold plasma is only in contact with a single electrode, the other one being covered by a dielectric barrier as sketched in Figure 2b. In this particular configuration, an auxiliary wire can be added into the capillary to propagate plasma over long distances. In 2018, it has been successfully applied and combined with pulsed electric fields technology to treat phosphate-buffered saline (PBS) so as to reduce the viability of lung fibroblasts and malignant melanoma cells [34], [35]. In 2020, Nascimento *et al.* have developed a flowing dielectric barrier device composed of an auxiliary floating electrode placed equidistant from 4 high voltage electrodes to generate a plasma jet [36]. The following year, the team upgraded its technology to a "transferred atmospheric pressure plasma jet" – like the one developed by Maho *et al.* to identify the OH and NO radicals resulting from a supply in helium mixed with molecular oxygen [34], [37]. In other APPJ configurations, the counter-electrode can be non-metallic and external to the APPJ device, as shown in Figure 2c. Although the electrical risks associated with this configuration are more difficult to control than in the previous configurations, this floating electrode plasma jet has been successfully utilized in dermatology to stimulate wound healing [38], [39].

An APPJ can be considered as a cold plasma catheter if it satisfies the following criteria: (i) its capillary is flexible, (ii) the maximum value of its outer diameter is lower than the inner diameter of the operating channel of a conventional endoscope, whether for duodenoscopy, gastroscopy, bronchoscopy or colonoscopy, i.e. typically lower than 4.2 mm and (iii) the thermal and electrical hazards are controlled whether the cold plasma catheter is used alone or placed inside a conventional endoscope. Technological miniaturizations and incrementations of the aforementioned APPJ devices have led to the development of cold plasma catheters that are now discussed:

(i) The plasma gun device (Figure 2d) is composed of an inner rod electrode and an outer ring electrode that are polarized and grounded respectively. One of its main features is that cold plasma is in direct contact with the polarized electrode. The guided streamers can propagate inside a flexible capillary, up to 60 cm in length if the voltage generator delivers pulses higher than 30 kV in magnitude with a width of a few microseconds [40]. This configuration offers the ability to propagate pulsed atmospheric plasma streams over longer distances by providing higher voltage values. Although this device has been successfully used as an APPJ in oncology applications [41], its utilization as a catheter remains limited if it is inserted into a hospital endoscope. Since the standard length of an endoscope is 2 meters, it would be necessary to apply voltages of the order of 100 kV for the propagation of plasma: a value that appears too high for the safety of both the practitioner and the patient.

(ii) The transferred plasma catheter (Figure 2e) is composed of an outer ring electrode polarized at high voltage, a grounded plate electrode and a long auxiliary wire distinct from these two electrodes [34], [42]. The ground electrode is far enough away from the polarized electrode (e.g. 20 cm) so that it does not influence the properties of the plasma generated in the device. An interesting characteristic of the transferred plasma catheter is that cold plasma is formed in three regions: (i) between the HV electrode and the proximal end of the auxiliary wire, (ii) along the auxiliary wire itself and (iii) from the auxiliary wire to the target (biological tissue, metal surface, …). Paradoxically, this auxiliary wire is both a strength and a limitation: if it allows the propagation of plasma over long distances, greater than 2 meters, the value of its voltage remains unknown during operation.

(iii) The bifilar helix electrode catheter (Figure 2f) is composed of a dielectric tube along which two wire electrodes (a polarized electrode and a grounded electrode) are wound helically and equidistant from each other. An external dielectric tube is coaxially superimposed on this arrangement to isolate the electrodes from the gaseous environment [43]. Such a catheter remains flexible and offers the possibility of delivering plasma several meters away within the capillary, as with the transferred plasma catheter. If the interelectrode space is greater than 2 mm, then a significant electric field is created inside the capillary and cold plasma is generated there. Interestingly, if this interelectrode space becomes lower than 2 mm, then narrow electrode winding forms a quasi-equipotential surface around the capillary, resulting in the absence of plasma generated within the capillary and therefore the advantage of not altering the inner walls of the capillary. The main drawback of this bifilar helix electrode is that a repeated utilization can modify the equidistant space separating the two wound electrodes and lead to plasma inhomogeneities. For this reason, the manufacture of this configuration requires resources and time.

(iv) The dual gas channel catheter (Figure 2g) comprises two flexible dielectric capillaries, coaxially centered and forming two separate gas channels. The polarized electrode is a metal wire that is twisted around the inner tube while a grounded electrode is placed at the distal end of the outer tube [44].

(v) The confined mono-electrode catheter (Figure 2h) is composed of a small-diameter dielectric capillary whose outer wall is covered by a metal deposit. Then, a dielectric capillary of larger diameter is coaxially placed all along the first one. The metal deposit is totally embedded by the two dielectric capillaries; it is biased to the high voltage and therefore plays the role of a polarized electrode [45]. This type of catheter may rise some concerns for endoscopic applications especially because it does not possess any grounded electrode while this requirement is a mandatory for the certification of the device either by the Food & Drug Administration (FDA) in United-States or by the Scientific Committee on Consumer Safety (SCCS) in European union.

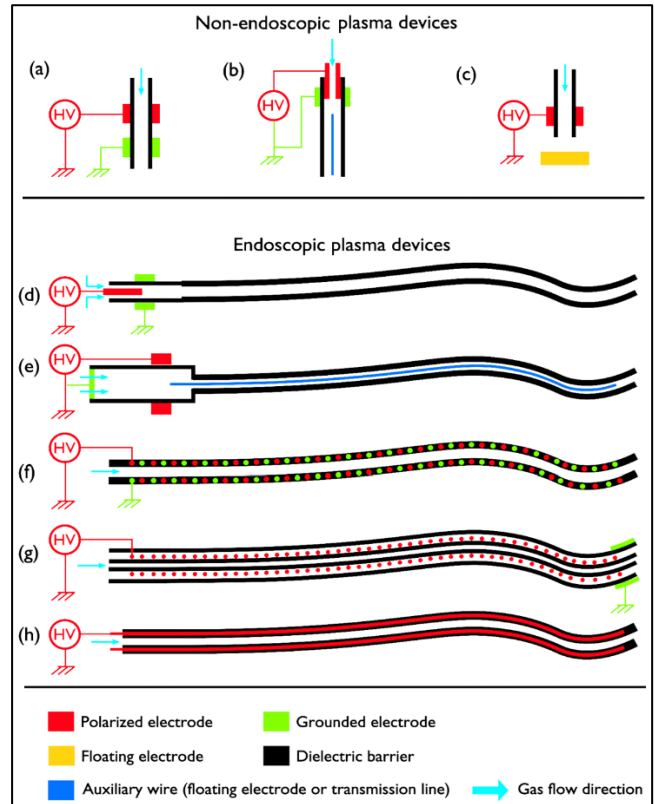

*Figure 2. Schematics of non-endoscopic and endoscopic plasma devices (a) plasma Tesla jet, (b) transferred atmospheric pressure plasma jet, (c) Floating electrode plasma jet, (d) plasma gun catheter, (e) transferred plasma catheter, (f) bifilar helix electrode catheter, (g) dual gas channel catheter, (h) confined electrode catheter.*

### I.2.5. Endoscopic utilization of cold plasma catheters

In the 1990s, cold plasma catheters were designed to deliver ionized argon gas onto injured tissues to induce blood coagulation [46]. Argon plasma is supplied by a high frequency current so that a high frequency electric field is confined between the electrodes of the applicator and the tissue. Since the energy has a penetration depth of 3 mm maximum, the process is both simple and risk-free [47]. The powerful but superficial heating of the tissue is responsible for closing the small blood vessels (coagulation) and for destroying the areas of pathological tissue (tissue devitalization). As an alternative to argon gas, carbon dioxide can also be supplied, as proposed by Kurosawa *et al.* for achieving hemostasis in cases of gastrointestinal bleeding [48].

Cold plasma catheters can also be considered for tumor resection although only non-endoscopic plasma jets have been utilized to date. Hence, a first proof of concept has shed light on the possibility of targeted ablation of bladder tumors grafted subcutaneously in nude mice [49]. Other works from Wang et al. show the ability of cold plasma approach for a selective ablation of metastatic BrCa cells *in vitro* without damaging healthy Mesenchymal Stem Cells (MSCs) at the metastatic bone site [50].

In addition to blood coagulation and tumor ablation, another challenge that cold plasma catheters could meet is the induction of antitumor effects through cell cycle arrest and cell apoptosis. Contrarily to the two previous applications, such approach does not rely on thermal effects but rather on a synergy resulting from plasma reactive chemical properties and electric field properties, the latter being able for example to induce cell oxidative stress and cell permeabilization respectively. This therapeutic strategy could be quite relevant to treat CCA, especially eCCA for at least two reasons. First because it has a much higher frequency (85-90%) than that of iCCA (10-15%). Second because it is implanted in the larger diameter bile ducts (cystic duct, bile duct, common hepatic duct) which facilitates the deposition of cold plasma on tumor sites.

No pre-clinical or clinical trial has been carried out so far to demonstrate antitumor effects obtained through standard endoscopic procedures. Therefore, and upstream of these objectives, it is necessarily to demonstrate the safety of a plasma catheter approach on models as close as possible to humans, for example on alive animals showing morphological similarities with human digestive system (e.g. anesthetized pig) or at least on human-like post-mortem anatomical models. This feasibility study can be achieved by demonstrating that cold plasma endoscopy does not imply any electrical or thermal hazards both for the patient and for the practitioner.

## II. Material & Methods

### II.1. Duodenoscope and cold plasma catheter

In these research works, a digital video duodenoscope from Pentax Medical Company (model ED-3480TK) is used, as sketched in Figure 3a. Thanks to its combination with Pentax DSP Technology and High Resolution colour CCD Chips, endoscopic images of brilliant quality are obtained regarding resolution and colour reproduction in the unique full screen size. Its insertion tube (11.6 mm of outer diameter and 125 cm of working length) possesses an operation channel with an inner diameter of 4.2 mm as highlighted in the cross-section view of Figure 3a. It is completed by an optical system with a field of view as high as 100° while its angulation is 120° up, 90° down, 100° right and 90° left [51].

Here, the transferred plasma catheter is selected to deliver cold plasma into biliary ducts models. This choice results from a benchmarking on the configurations presented in Figure 2 and which is based on the following motivations:

(i) <u>Plasma gun catheter</u>: To propagate the cold plasma all along the insertion tube of a conventional duodenoscope, i.e. 2 meters long, it would be necessary to use a generator capable of delivering voltages higher than 30 kV. However, our nanoimpulse generator is limited to 10 kV and cannot propagate the plasma over distances greater than 60 cm in this catheter configuration.

(ii) <u>Transferred plasma catheter</u>: Its design is based on technical means that are easy and quick to implement. This catheter allows the cold plasma to be propagated over distances of up to 5 meters only if a metal auxiliary wire runs along the capillary.

(iii) <u>Bifilar helix electrode catheter</u>: This configuration is the most difficult to implement for at least two technical reasons. First, it requires to wind two wire electrodes in the form of a double helix while maintaining a regular and precise inter-electrode gap (< 0.5 mm). Second, it is mandatory to cover these two wire electrodes by an external dielectric layer to prevent any electrical contact with biological tissues or with the internal walls of the duodenoscope. Other constraints apply to these specifications, notably the flexibility that the catheter must maintain to facilitate its passage through the working channel of duodenoscope as well as the inter-electrode distances that must remain as equidistant as possible. These constraints are all the more challenging to satisfy that the catheter is submitted to twisting and bending but also to repeated use (except if the catheter is intended for single use).

(iv) <u>Confined electrode catheter</u>: this configuration, at first sight aesthetically simple, requires the ability to isolate the polarized electrode by a double dielectric barrier (internal and external) like the previous configuration. This device does not have a ground electrode, hence limiting further commercialization for clinical purposes since all medical electrical devices must meet this criterion.

As sketched in Figure 3b, this plasma source is composed of two organs: (i) a main chamber supplied with helium gas at a flow rate of 1 slm in which cold plasma is primarily generated and (ii) a long flexible capillary which allows the plasma to be propagated over long distances. This capillary is made of polytetrafluoroethylene (PTFE), has an internal diameter of 1 mm, a thickness of 0.5 mm and a length of 2 m. The main chamber has two electrodes: a flat electrode connected to the nanopulse high voltage generator device (model Nanogen 1 from RLC Electronic company) and an outer ring electrode that is grounded. A 1 mm thick dielectric barrier covers the polarized electrode to prevent any passage to the thermal arc regime. An auxiliary copper wire (0.3 mm in diameter) is coaxially centered in the main chamber and runs along the flexible capillary. Its length is 0.5 cm shorter than the length of the flexible capillary so that it remains confined in the capillary and cannot be in contact with living tissue. In this configuration, cold plasma is formed between the polarized electrode and the auxiliary wire, then propagates all along the auxiliary electrode and finally exits the capillary as a plume of several millimeters long.

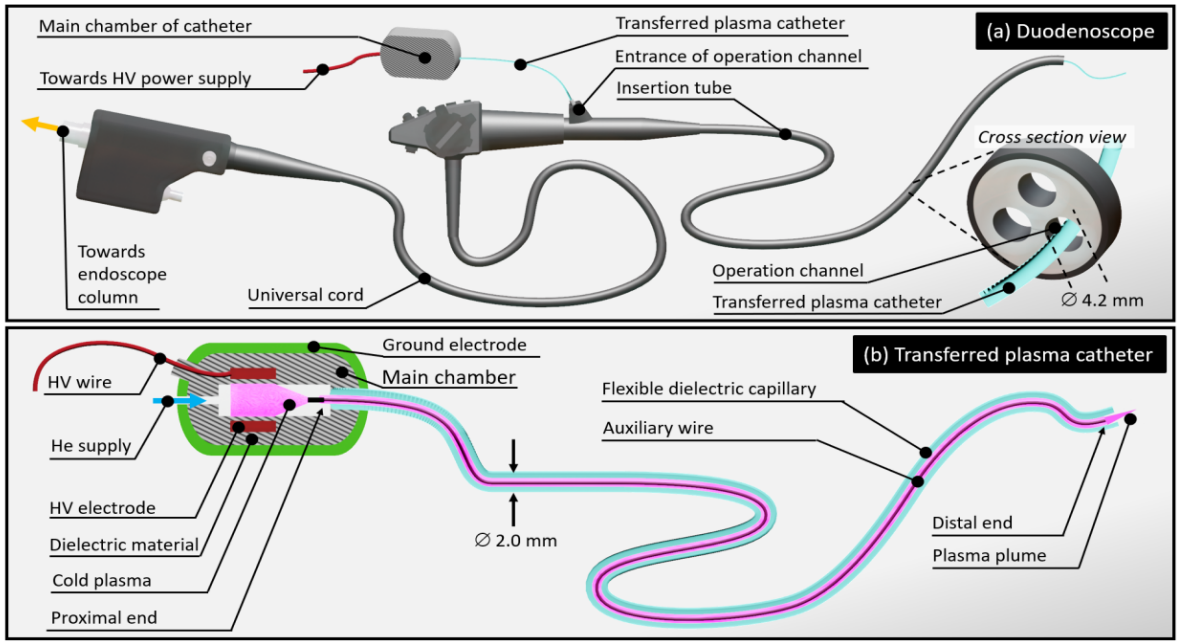

*Figure 3. (a) Schematics of a conventional duodenoscope with a cold plasma catheter inserted in the operation channel, (b) Schematics of the transferred plasma catheter.*

### II.2. Biliary tree models

To validate the relevance of cold plasma endoscopy to bile ducts diseases, experiments are successively carried out on two models of biliary trees: an artificial human biliary tree model (AHBT) and a post-mortem porcine biliary tree model (PPBT). All procedures with an endoscope have been performed by MC, endoscopist.

### III.2.1. The AHBT model

The artificial human biliary tree model (AHBT) model is made from two elements:
(i) A biliary endoscopy trainer from Chamberlaingroup company (model 2101) utilized in routine by apprentices to simulate endoscopic retrograde cholangiopancreatography (ERCP) on fictive patients placed either in lateral or prone position. Access is gained through a long tube that simulates passage of a scope through the esophagus, the stomach and the duodenum. The trainer also includes a port for fluid injection to enhance the realism of the training experience. As shown in Figure 4a, the biliary tree features the lumen of the common bile duct, cystic duct, hepatic ducts and pancreatic duct. It can be confined from ambient air by a transparent protective cover. Furthermore, it can be completed by stricture inserts and/or tumor volume inserts (Figure 4b) to represent the diversity of the pathological cases encountered at hospital.
(ii) A tumor insert exhibiting the same electrical response as a real tumor. For this, a thin aluminum layer is deposited on a tumor volume insert, as shown in inset of Figure 4a. Then, it is connected to a resistor and a capacitor in parallel, the whole being grounded (Figure 4b). Their values ($R_{eq}$ = 1.5 kΩ, $C_{eq}$ = 100 pF) are chosen to mimic the electrical parameters of the human body and are based on our previous research works [52].

The Figure 4c is a photograph of the resulting AHBT model where cold plasma is generated at the distal end of the common bile duct.

### III.2.2. The PPBT model

Three specimens of the post-mortem porcine biliary tree model (PPBT) are purchased at LEBEAU company, specialized in the transport and trade of animal anatomical parts for research and education. As shown in Figure 5, the model is composed of esophagus, stomach, duodenum, biliary tree, liver and gallbladder. The PPBT is placed in a dedicated plastic structure shape and size resembling the human peritoneal cavity. This structure, known as EASIE (Erlangen active simulator for interventional endoscopy) has been utilized under its EASIE-R$^{TM}$ version from ENDOSIM company. EASIE-R$^{TM}$ is an excellent educational tool for interventional endoscopy with a high level of acceptance, whether for endoscopic ultrasound procedure (EUS), natural orifice translumenal endoscopic surgery (NOTES) or cold plasma endoscopy (CPE) as demonstrated in this article. All experiments on PPBT models have been carried out at the surgical school of AP-HP, Fer à Moulin, Paris.

### III.3. Plasma diagnostics

### II.3.1. Electrical parameters

The electrical parameters of cold plasma are monitored using an analog oscilloscope (Wavesurfer 3054, Teledyne LeCroy). Voltages are measured with a high-voltage probes (Tektronix P6015A 1000:1, Teledyne LeCroy PPE 20 kV 1000:1, Teledyne LeCroy PP020 10:1) while current peaks are analyzed using a current transformer (Pearson company, model 2877). Electromagnetic radiation from the high voltage generator is removed from the currents and voltages measured using a Butterworth filter [53].

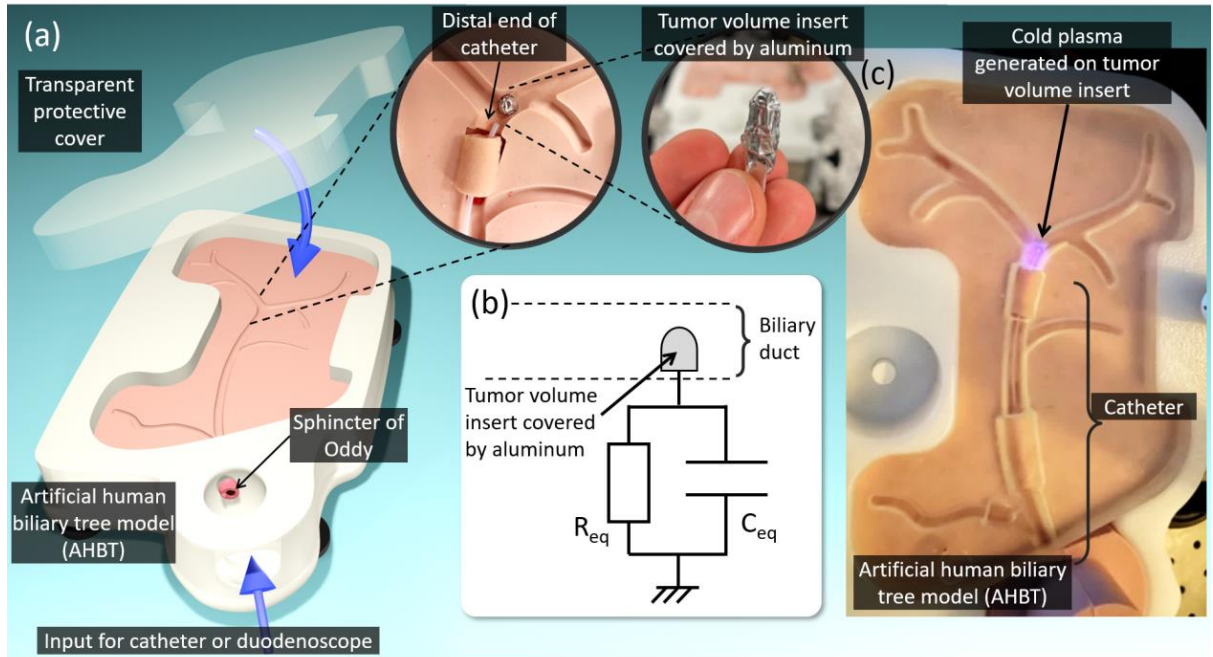

*Figure 4. (a) Diagram of the ERCP trainer with enlarged photo of the distal end of the catheter and enlarged photo of the tumor volume insert, (b) Equivalent electrical model of tumor, (c) Transferred plasma catheter in operation inside the ERCP trainer. Cold plasma is observed through the transparent protective cover at the extremity of the common bile duct.*

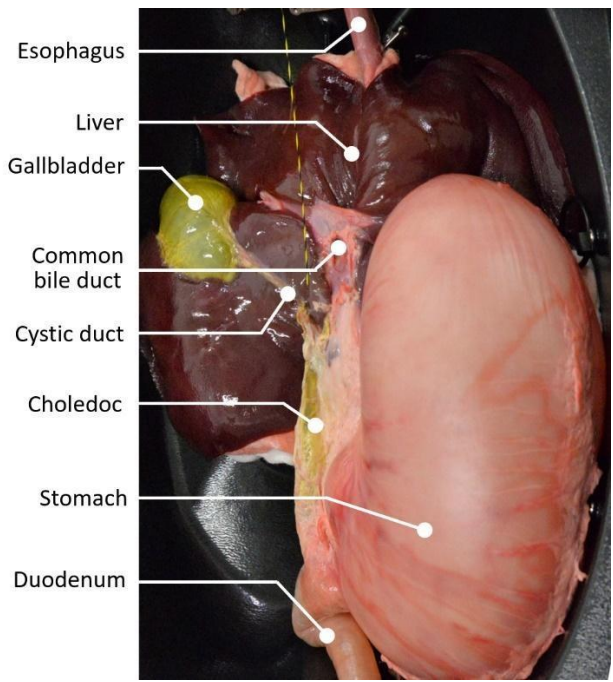

*Figure 5. Anatomical description of the post-mortem porcine biliary tree model (PPBT) with adjacent organs (esophagus, stomach, liver, duodenum) placed in EASIE-R<sup>TM</sup>.*

### II.3.2. Radiative species

Optical emission spectroscopy is utilized to identify the radiative species from plasma. The spectrometer (SR-750-B1-R model from Andor company) is equipped with an ICCD camera (Istar model) which operates in the Czerny Turner configuration. Its focal length is 750 mm while diffraction is achieved with a 1200 grooves.mm$^{-1}$ grating in the visible range. The following parameters are selected for all experiments: exposure time = 0.1 s, number of accumulations = 50, readout rate = 3 MHz, gate mode = 'CW ON', intensification factor = 2000, insertion delay = 'ultra fast'.

### II.3.3. Temperature measurements

Temperature measurements of material surfaces or biological tissues exposed to plasma are performed using thermographic infrared imaging. The camera from JENOPTIK & INFRATEC is equipped with a VarioCAM® HD head 680/30 and allows real-time analysis, transfer and processing of information. Its spectral range is 7.5-14 μm, its thermal measurement range is -40°C to 1200°C, its thermal resolution at 30°C is down to 30 mK, and its measurement accuracy is ± 1,5% from 0°C to 100°C.

## II.4. Biological diagnostics

### II.4.1. Cell culture and cold atmospheric plasma (CAP) treatment

HuCCT-1 cells, derived from intrahepatic biliary tract, were kindly provided by Dr. G. Gores (Mayo Clinic, MN). EGI-1 cells, derived from extrahepatic biliary tract, were obtained from the German Collection of Microorganisms and Cell Cultures (DSMZ, Germany). Cells were cultured in DMEM supplemented with 1 g/L glucose, 10 mmol/L HEPES, 10% fetal bovine serum (FBS), antibiotics (100 UI/mL penicillin and 100 mg/mL streptomycin), and antimycotic (0.25 mg/mL amphotericin B). Cell lines were routinely screened for the presence of mycoplasma and authenticated for polymorphic markers to prevent cross-contamination.

The cells were treated directly with the catheter described in Figure 3b. It was supplied with helium gas (Alphagaz 1 standard from Air Liquide company) at a flow rate of 1 slm and powered with a nanopulse high voltage generator device (model Nanogen 1) from RLC Electronic company. Electrical parameters were fixed as follows: 8 kV of amplitude, 1% of duty cycle and 5 kHz of repetition frequency.

### II.4.2. Cell viability

20 000 cells/well were plated in 24-well plates. 24 h later, the medium was replaced by fresh culture medium and the cells were exposed directly to plasma during 0.5, 1, 2, 3 and 5 min. Cells were then incubated for 72 hours before determining the viability by the crystal violet method. Absorbance was quantified with a spectrophotometer (Tecan) at 595 nm.

## III. RESULTS & DISCUSSION

This section starts with a physical characterization of the transferred plasma catheter to explain how cold plasma is successfully propagated over long distances despite voltages of a few kV. In a second step, the catheter is applied directly to the AHBT model, without being inserted into the duodenoscope. This study aims at validating the absence of electrical and thermal risks likely to invalidate the selected configuration. In a third step, the transferred plasma catheter is introduced into the operating channel of the duodenoscope, the whole being inserted into the PPBT model. This anatomical porcine model, closer to a real patient case, is the subject of a new feasibility-risk study. In a fourth step, the therapeutic efficacy of cold plasma endoscopy is demonstrated on CCA cells as part of an *in vitro* experimental campaign. As a reminder, the antitumor efficacy is not demonstrated in alive porcine models, as such approach does not exist in the case of CCA.

### III.1. Deciphering the role of the auxiliary wire in the transferred plasma catheter

Of the different configurations shown in Figure 2, only the transferred plasma catheter (e) has a long auxiliary wire running along its capillary, one end being in contact with the plasma in the main chamber and the other with the plasma plume. It is worth stressing that the auxiliary wire is not an option but a mandatory to propagate cold plasma over long distances. Without the auxiliary wire, the guided streamers can only propagate less than 10 cm along the 2 m long flexible capillary. Figure 6a shows the voltage profiles measured at the proximal and distal ends of the auxiliary wire during plasma operation ($V_{Generator}$ = 10 kV, $D_{Cycle}$ = 1%, $f_{rep}$ = 10 kHz, $\Phi_{He}$ = 1 slm). The amplitudes are quite different with values as high as 1400 V and as low as 350 V at the proximal and distal ends respectively. The current, however, always has the same profile whether at proximal end, distal end (data not shown) or mid-length. The time profile measured at mid-length is shown in Figure 6b: two peaks of conduction current are measured at the rising edge (t = 5 µs) and falling edge (t = 7 µs) of the high voltage pulse, corresponding to positive and negative guided streamers respectively, as already evidenced in [53]. Consequently, the time profiles of $V_{wire}$ and $I_{wire}$ enable to assess the electrical power in the auxiliary wire: its value from the proximal end to the distal end decreases from 0.67 W to 0.25 W, a 63% loss. In addition, the impedance of the auxiliary wire ($Z_{wire}$) can be obtained using equation {1}. For the sake of clarity, the Figure 6c represents $1/Z_{wire}$ as a function of time. In absence of cold plasma (i.e. no guided streamer and no peak of conduction current), the impedance tends to an infinite value, hence explaining why $1/Z_{wire}$ is close to zero before 5 µs, between 5.1 and 7 µs, and after 7.1 µs. The notion of impedance appears during the lifespan of the two conduction peaks, with a value close to 11 kΩ at the rising edge of the pulse (5-5.1µs) and 6 kΩ at the falling edge (7-7.1µs). These values should be considered as rough estimates due to uncertainties in the three measured parameters ($V_{wire,pr}$, $V_{wire,dist}$ and $I_{wire,mid}$) but overall, $Z_{wire}$ can be estimated to be on the order of 10 kΩ.

$$Z_{wire} = \frac{V_{wire,pr} - V_{wire,dist}}{I_{wire,mid}} \quad \{1\}$$

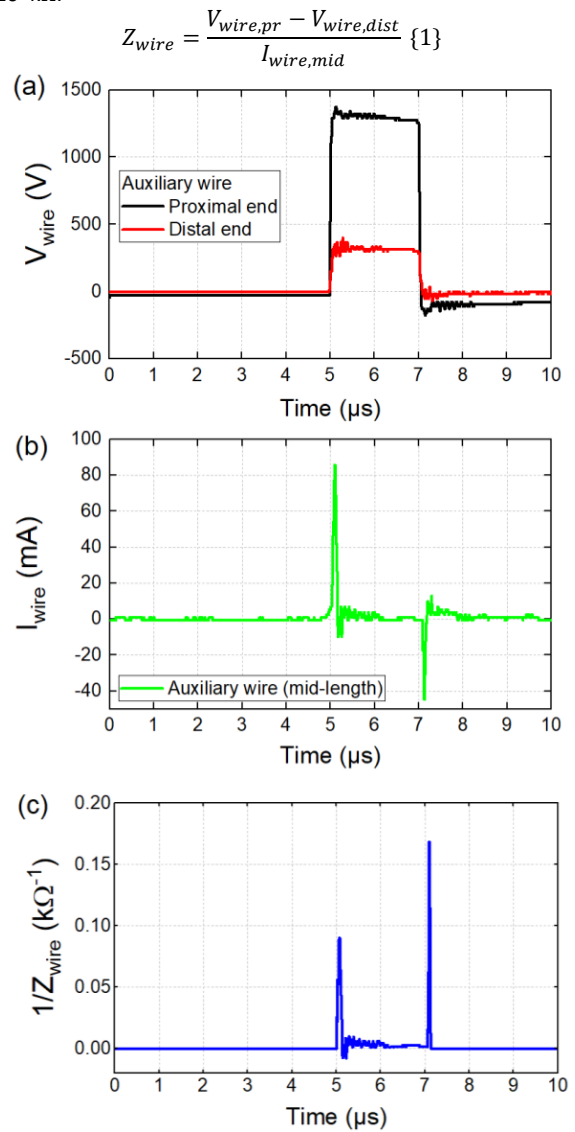

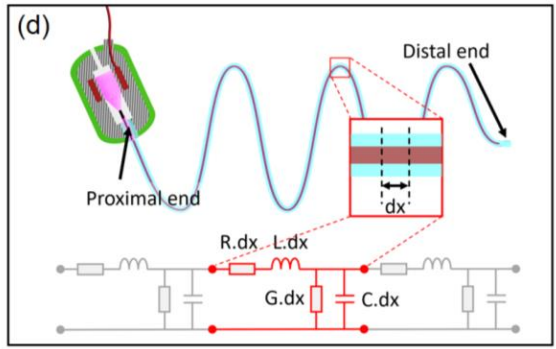

Figure 6. (a) Time profile of the voltage measured at proximal and distal ends of the transmission line, (b) Time profile of the current measured at mid-length of the same electrode. In both cases, the transferred plasma catheter is not inserted into the duodenoscope ($D_{Cycle}$ = 1%, $f_{rep}$ = 5 kHz, He flow rate = 1 slm). (c) Impedance of the auxiliary wire ,(d) Schematic representation of the elementary components of a transmission line.

The understanding of these results is based on the principle that the auxiliary wire can behave either as a floating electrode or as a transmission line depending on whether its length ($L_{wire}$) is negligible compared to the spatial period of the voltage signal which traverses it. In this work, since the excited electrode of the catheter is polarized by high voltage pulses that are regularly repeated, cold plasma is generated as guided streamers. The spatial period ($\lambda$) associated with a train of guided streamers can be defined as the ratio of the streamer propagation velocity (v) by its temporal repetition frequency ($f_{rep}$), as indicated in relation {2}:

$$\lambda = \frac{v}{f_{rep}} \quad \{2\}$$

In our previous research works performed on atmospheric pressure plasma jets, fast ICCD imaging shows that the guided streamers can propagate from 100 km.s$^{-1}$ to more than 2000 km.s$^{-1}$ depending on the position where velocity is measured [53]. Here, the mean propagation velocity measured by fast ICCD imaging within the catheter is approximately 100 km.s$^{-1}$. Since the signal frequency is set at 10 kHz, $\lambda$ is estimated to be close to $100.10^3$ [m/s] /$10.10^3$ [Hz] = 10 m. This value is probably overestimated because the Fourier signal decomposition of the pulsed signal shows the presence of harmonics at much higher frequencies due to the very fast rising edge of the voltage signal (30 ns). Consequently, if $L_{wire}$ << 10 m (i.e. a few centimeters) then the auxiliary wire behaves like a floating electrode: the electric potential is uniform at all points and invariable over time. However, if $L_{wire}$ has a value close to 10 m or that is not negligible compared to 10 m, then the auxiliary wire behaves like a transmission line (or waveguide). Here, since $L_{wire}$ = 2 m, the auxiliary wire can be modeled as a transmission line, as already suggested by Bastin *et al.* [42]. In order to understand these results, the catheter is modeled as an infinite series of two-port elementary components, each representing an infinitesimally short segment of the transmission line whose cross-section is considered constant along its length. These two-port elementary components are detailed in Table 2 and a schematic of their arrangement is proposed in the equivalent electrical model in Figure 6d. The series inductors can be neglected because the voltage profile from Figure 6a and the current profile from Figure 6b do not show significant phase shift. However, shunt resistors must be considered. Indeed, when the catheter is powered, plasma can be observed non-uniformly along the flexible capillary, as photographed in Figure 9b. Thus, the emissive regions which indicate the presence of plasma, can be modeled by a resistor, while the non-emissive regions indicate the presence of non-ionized helium and can therefore be modeled by a capacitor. The catheter behaves like a transmission line characterized by significant energy losses resulting from the non-uniformly propagating plasma regions all along. Finally, it is worth stressing that the terminology of "floating electrode" is inappropriate in our study for at least two reasons: first the auxiliary wire is traversed by a potential gradient (while by definition, voltage is uniform within a same electrode) and its impedance is 80 kΩ (while by definition, it must be zero in any electrode).

| Measurands | Components | Symbols | Units |
|---|---|---|---|
| Resistance of the conductors along the line | Series resistor | R | $\Omega.m^{-1}$ |
| Inductance along the line resulting from the magnetic field associated to the streamers propagation | Series inductor | L | $H.m^{-1}$ |
| Capacitance between auxiliary wire and outer walls of the catheter | Shunt capacitor | C | $F.m^{-1}$ |
| Conductance of the gas between auxiliary wire and catheter's walls | Shunt resistor | G | $S.m^{-1}$ |

Table 2. Elementary components of a transmission line.

## III.2. Transferred plasma catheter applied to the artificial human biliary tree (AHBT) model

### III.2.1. Electrical properties (catheter operating outside the duodenoscope)

The transferred plasma catheter is placed in the AHBT model so that the distance between its distal end and the equivalent electrical tumor insert is 10 mm (i.e. gap = 10 mm). The time profiles of voltage and current deposited on the equivalent electrical tumor insert are shown in Figures 7a and Figure 7b respectively. It turns out that the maximum values are approximately 40 V and 10 mA respectively. Considering that the signals are close to pulse profiles, their RMS values can be deduced from relation {3} and correspond only to 3.98 V and 1.19 mA respectively. Then, the electrical power dissipated in the equivalent electrical tumor insert is assessed using relation {4}:

$$V_{RMS} = V_{max} \cdot \sqrt{D_{cycle}} \quad \{3\}$$

$$P_{Deposited} = f_{rep} \cdot \int_T I_{Deposited}(t) \cdot V_{Deposited}(t) \cdot dt \quad \{4\}$$

As shown in Figure 7c, the influence of the generator voltage from 5.5 kV to 8.0 kV as well as the influence of the repetition frequency from 1 to 10 kHz drive to power values always lower than 5 mW, which – to a first approximation – corresponds to power densities close to 10 mW/cm$^3$ in the biliary ducts. Interestingly, the Figure 7d indicates that this electrical power has a non-linear dependence with the gap. While $P_{Deposited}$ remains always lower than 2 mW for gaps higher than 3 mm, $P_{Deposited}$ reaches a value as high as 40 mW for a 3mm-gap. For this reason, the gap is a strategic parameter that needs to be finely controlled in interventions such as ERCP. By playing on it, it is possible to control the deposited power and therefore the temperature of the exposed tissues, as discussed in the next section. The same question may arise for another parameter of the catheter: its flexibility. Since the catheter is utilized into the AHBT model, it is already bent by passing through the anatomical structures mimicking the duodenum and the choledoc of the biliary tree (see Figure 4). However, to verify that its electrical response would remain unchanged to treat other parts of human body, the catheter has been utilized outside the AHBT model and bent into different configurations depicted in Figure 7.e, where each configuration is characterized by a set of bending angles (α = 0°, 5°, 15°, 25°, 35° and 180°). For a gap fixed at 5 mm, the Figure 7.f indicates that the deposited power from relation {4} remains roughly the same in all the aforementioned configurations, with values lower than 3 mW. Finally, one must stress that these profiles are obtained when the catheter is outside the duodenoscope. In section III.3, we will see how the "catheter in duodenoscope" configuration significantly modifies the electrical characteristics of plasma.

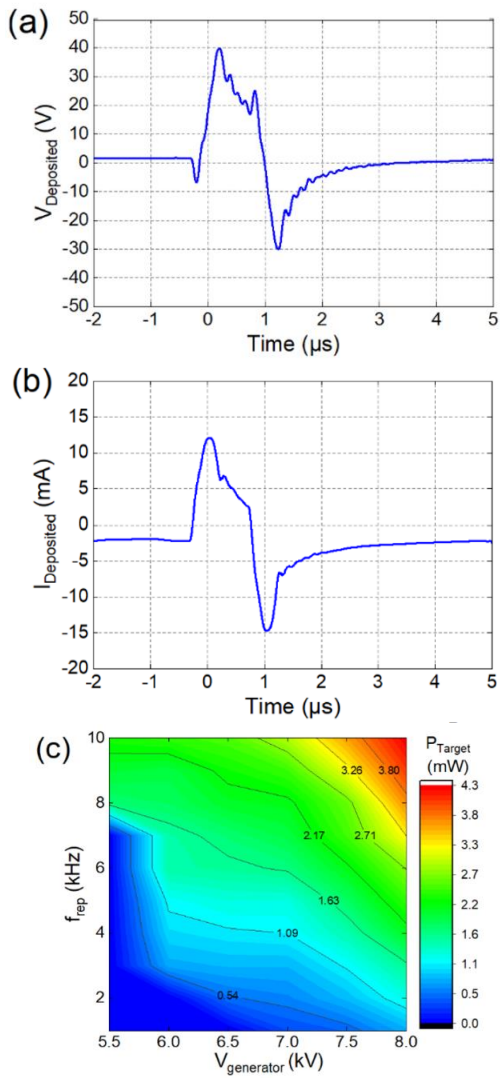

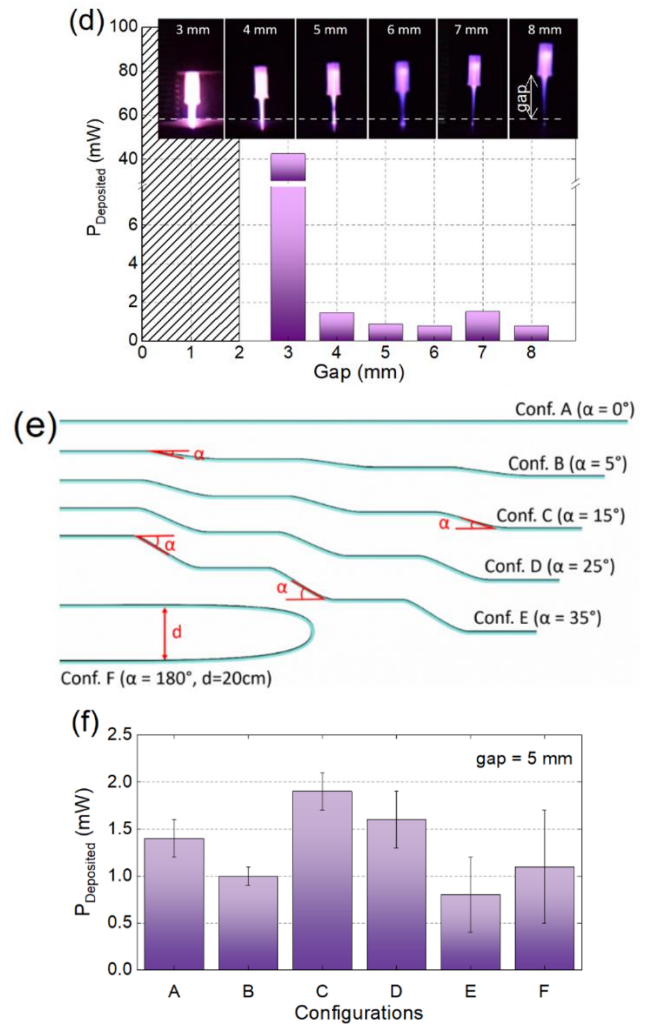

*Figure 7. (a) Time profile of the voltage measured on the AHBT tumor insert, (b) time profile of the current measured on AHBT tumor insert, (c) power dissipated in the AHBT tumor insert as a function of the voltage delivered by the generator and its repetition frequency, (d) power measured in the AHBT tumor insert versus the gap, (e) Bending configurations of the catheter outside AHBT model, (f) Influence of bending configurations on the electrical power deposited by the catheter on target. In all cases, the transferred plasma catheter is not inserted in duodenoscope ($D_{Cycle}$ = 1%, $f_{rep}$ = 10 kHz, $\Phi_{He} = 1\ slm$, gap = 10 mm).*

### III.2.2. Thermal properties (catheter operating outside the duodenoscope)

Figure 8 shows infrared photographs of a portion of the AHBT model where the transferred plasma catheter is in operation. The photographs are taken at 1 min and 10 min after cold plasma ignition, without/with the transparent protective cover (as sketched in Figure 4a). Several observations are noteworthy:

(i) The temperature values are entirely compatible with the targeted medical application since the 40°C threshold is never exceeded.

(ii) Inside the bile ducts, the temperature goes from about 22 °C to 24 °C after 1 min to stabilize at a maximum value of 28°C after 10 minutes of operation.

(iii) In Figure 8a, the gaseous effluent propagates preferen-tially in the "right bile duct" whose internal walls are heated by the plasma plume. Nevertheless, this same effluent also propagates isotropically over shorter distances without being ionized, leading to a very localized cooling of the internal walls but still significant since a value of approximately -1.0 °C is obtained (the infrared camera having a thermal sensitivity of 0.05 °C).

(iv) The addition of the protective cover (1 cm thick) qualitatively illustrates the diffusion of heat emitted by plasma in adjacent tissues. The temperature increases from 22 °C to 24 °C after 1 min exposure, to reach 28 °C after 10 min. Maximum temperatures obtained in Figure 8c and 8d do not drive significant discrepancies.

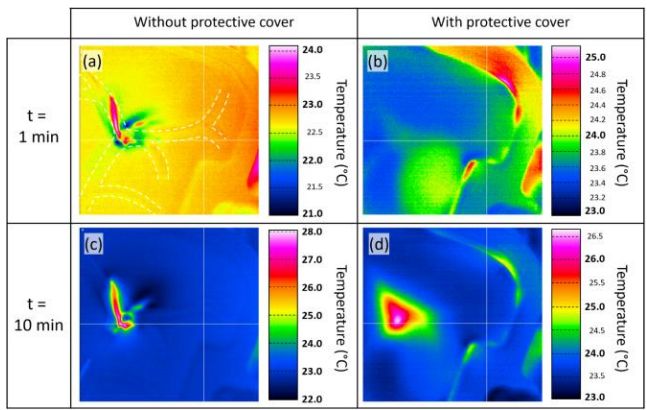

*Figure 8. Infrared pictures of the AHBT model where cold plasma is delivered in the right bile duct. The transferred plasma catheter operates at $V_{Generator}$ = 7 kV, 10 kHz, $\Phi_{He} = 1\ slm$. Pictures are taken at t = 1 min and t = 10 min after plasma exposure, with/without transparent protective cover.*

# III.3. Transferred plasma catheter in duodenoscope applied to the post-mortem porcine biliary tree (PPBT) model

## III.3.1. Experimental framework

The technical feasibility of the transferred plasma catheter has been demonstrated on a model reproducing the topography of the bile ducts while mastering the electrical and thermal risks. Now, we propose to use a model closer to the human, in the perspective of carrying out clinical research in the near future. The photograph in Figure 9a shows the experimental environment in which this new study is carried out: the PPBT model (whose anatomy is detailed in Figure 5) is placed in the EASIE-R[TM] on an X-ray table. The duodenoscope is connected to the interventional endoscopy column which manages the incoming/outgoing fluids, the light source of the duodenoscope and the real-time video system (for data acquisition and processing ). An X-ray imager is also used to perform real-time radiographic monitoring of the endoscopic tools (catheter, duodenoscope, instruments, etc.) in the cavities of the PPBT model.

Before inserting the transferred plasma catheter into the duodenoscope to treat the bile ducts, a preliminary maneuver is carried out percutaneously, i.e. without using the duodenoscope. For this, a thin surgical incision is made in the duodenum to introduce the catheter in the bile ducts. Figure 9b is a night-time photograph which shows that the part of the catheter left outside the porcine model, shows both emissive regions (indicating the presence of plasma) and dark regions (indicating the presence of un-ionized helium). Since the electrical and thermal parameters have shown values close to those obtained in the previous section, a conventional endoscopic intervention can now be carried out following a 5 steps-sequence:

(i) The transferred plasma catheter is introduced into the working channel of the duodenoscope which remains outside the PPBT model. The catheter is supplied at 10 kV to ensure that a plasma plume can be generated. The "catheter + duodenoscope" configuration is validated, as evidenced in Figure 9c since the plume is visible and can reach a length of approximately 1 cm. Then, plasma is switched off.

(ii) The duodenoscope is inserted into the esophagus, the stomach and then the duodenum to reach the papilla which marks the entrance to the biliary tree.

(iii) Using a guide wire (not to be confused with the auxiliary wire), the endoscopist slides the plasma transferred catheter until it passes through the papilla. The catheter is introduced into the common bile duct of the biliary tree to ascend to the common bile duct.

(iv) The catheter is supplied with helium at a flow rate of 1 slm. This value is neither a low nor a high limitation since the catheter can generate cold plasma for values between 0.2 and 10 slm. Note, however, that there is a threshold limit of 8 slm that should not be exceeded to prevent any risk of biliary barotrauma or fatal gas embolism, as underlined by Mukewar et al. who injected $CO_2$ into the biliary tree of porcine models [54].

(v) Plasma in the catheter is switched on for about twenty minutes during which electrical and thermal measurements are performed. Figure 9c is a night photograph of the PPBT model that shows cold plasma illuminating the bile duct from within.

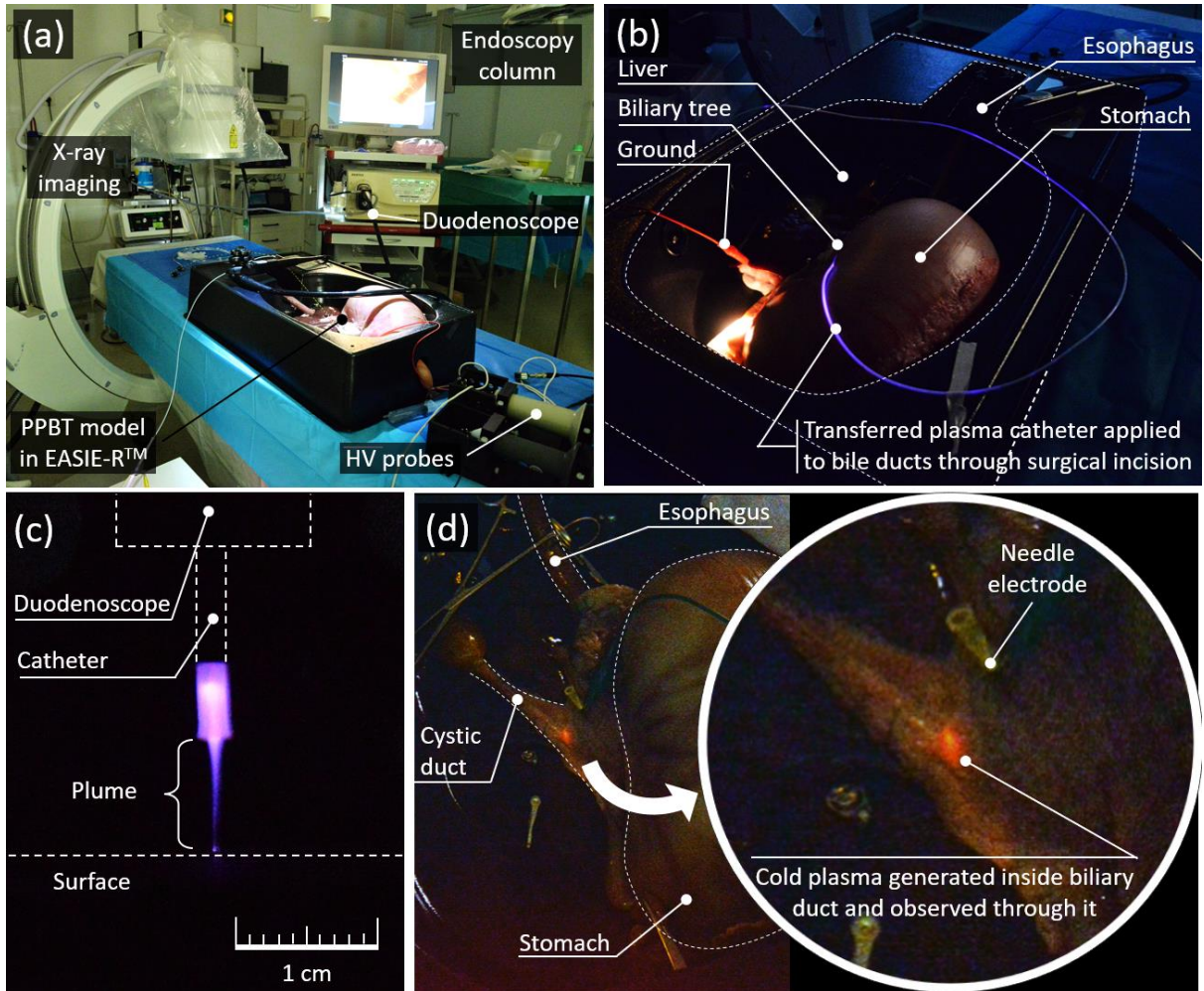

*Figure 9. (a) Photograph of the operation room where a post-mortem porcine biliary tree model is contained in EASIE-R[TM] with endoscopy and plasma equipments, (b) Night vision of the PPBT model in which a fine surgical incision is achieved in the duodenum to insert the transferred plasma catheter directly into the bile duct without having to use the duodenoscope, (c) Night vision of the plasma plume emitted by the transferred plasma catheter inserted in the duodenoscope (d) Night vision of the PPBT model where cold plasma is delivered in the biliary duct and observed through it.*

### III.3.2. Electrical properties

To verify the absence of electrical hazards inherent in the cold plasma endoscopy procedure, the current, voltage and power deposited in the bile ducts are measured using the setup shown in Figure 10a. When the cold plasma plume is generated in the common bile duct at point A, a needle electrode is placed in adjacent tissue approximately 20 mm away (point B). This electrode is then connected to the equivalent electrical model, already used in section III.2. (1.5 kΩ resistor in parallel with a 100 pF capacitor). Figure 10b shows the $P_{Model}$ values measured for $V_{Generator}$ between 8 and 10 kV and $f_{rep}$ set at 5 and 10 kHz. Through these 5 experimental conditions, it appears that the electrical power typically varies between 250 and 500 mW: values that are higher than those obtained when the catheter operates outside the duodenoscope but which lead to reasonable heating, as we will see in the thermal study of section III.3.3. Ahead of these electrical power values, it is important to look at the temporal profiles of voltage and current dissipated in the PPBT model. While Figure 10c recalls the temporal profile of the voltage delivered by the generator ($V_{Generator}$ = 10 kV, $f_{rep}$ = 10 kHz, $D_{Cycle}$ = 1%), Figures 10d and 10e represent the temporal profiles of $V_{Deposited}$ and $I_{Deposited}$. The voltage and current reach maximum values of approximately 200 V and 1 A respectively and RMS values of 20 V and 100 mA respectively. The RMS value of current is overestimated since the current peaks do not have a constant 1A-magnitude during the 1 μs pulse width. A more accurate calculation shows an RMS value of 32 mA. Although it is common practice to assess electrical risks through these RMS parameters, their use here is irrelevant as detailed afterwards (Figure 13) at the light of the standards ruling the safety of medical devices delivering short-time current pulses.

At this stage, three questions legitimately arise: why is the absolute magnitude of current multiplied by a factor of 1/0.012 = 80 (depending on whether the catheter is inserted into the duodenoscope or not)? What types of undesirable electromagnetic effects can result from the nanopulsed excitation of cold plasmas? Are the electrical parameters safe for a patient undergoing such a plasma endoscopic procedure? The answers to these questions are discussed below.

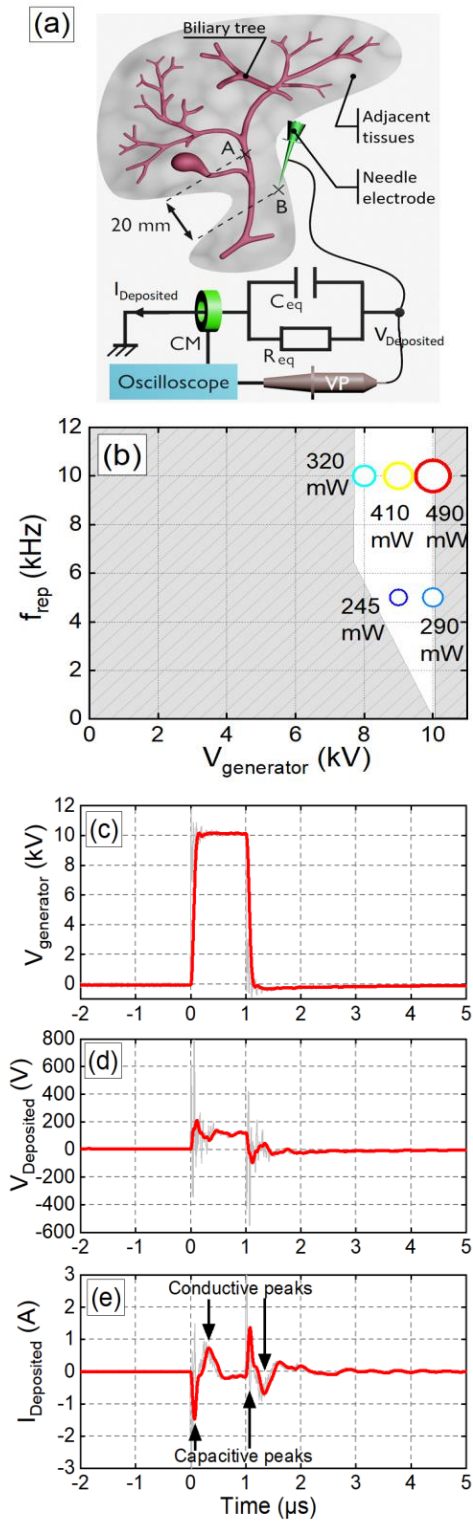

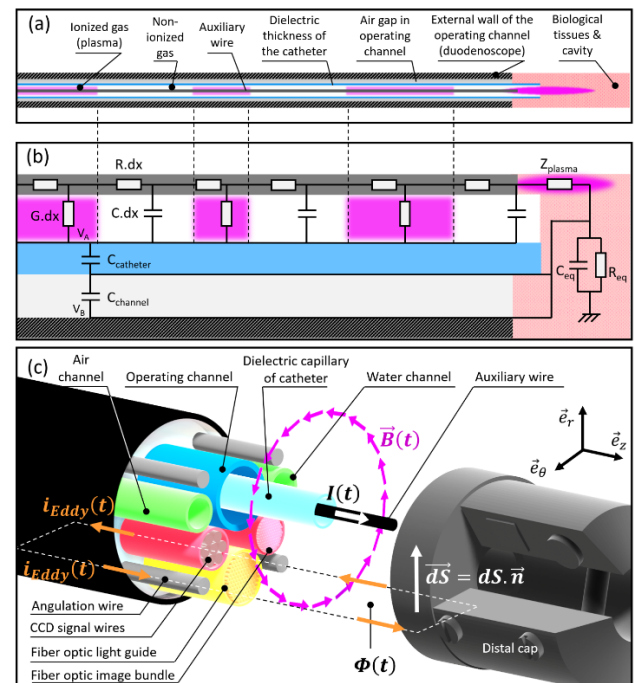

**(a) Why is the current increased if the catheter is inserted into the duodenoscope?**

First, the Figure 11a recalls that the gas confined in the dielectric capillary is only ionized at some places. The auxiliary wire behaves as a transmission line whose elementary conductances (G.dx) and capacitances (C.dx) stand respectively for these ionized and non-ionized gas regions, as shown in Figure 11b. This modeling is completed with $C_{Catheter}$ and $C_{Channel}$ which correspond to the capacitance of the catheter (more precisely the dielectric thickness of its capillary) and the capacitance of the gaseous environment within the working channel respectively. This set is electrically connected to the biological tissues modeled by the paralleling of $R_{eq}$ and $C_{eq}$. When the catheter is operating outside the duodenoscope, the $V_B$ potential (Figure 11b) is floating, so the leakage currents passing through $C_{Catheter}$ are relatively small. Since they cannot contribute to a charge transfer to the plasma pen, they can also be referred to as "loss currents." In contrast, when the catheter is in the duodenoscope, the $V_B$ potential is no longer floating. Indeed, the catheter is inserted in the operating channel of the duodenoscope, the internal walls of which are at certain sections metallic and connected to the ground. Consequently, $V_B$ is grounded, which drives to ($V_A - V_B$) values larger than those obtained in the simple catheter configuration, and therefore in much higher leakage currents. These leakage currents are not lost: they are transferred to the tissues to be treated by simple contact with the duodenoscope. These leakage currents can thus contribute to significantly increasing the value of $I_{Deposited}$, as shown in Figure 10e.

*Figure 10. (a) Sketch diagram showing how voltage and current are measured in porcine body, (b) Electrical power deposited by cold plasma in the biliary ducts versus voltage magnitude and voltage frequency ($D_{Cycle}$ = 1%, $\Phi_{He}$ = 1 slm), (c) Temporal profile of the applied voltage for a magnitude of 10 kV, $D_{Cycle}$ = 1% and $f_{rep}$ = 10 kHz, (d) Temporal profile of the voltage measured in PPBT, (e) Temporal profile of the current measured in PPBT.*

*Figure 11. (a) Diagram of the transferred plasma catheter interacting with biological tissues, (b) Equivalent electrical model of the "catheter in duodenoscope in biliary duct" assembly, (c) Diagram explaining the generation of eddy currents in the aforementioned assembly.*

**(b) Undesirable electromagnetic effects**

Electromagnetic radiation (EMR) is emitted by the HV generator during the endoscopic procedure, creating interferences on the communication and display systems (CCD camera embedded in the duodenoscope, monitors of the endoscopic column, ...). The images transmitted in real time are altered by the appearance of regularly spaced parasitic patterns, which can interfere with the analysis work of endoscopists, especially when treating patients with pathological or anatomical complications (e.g. stenosis). The Figure 12 shows a fast Fourier transform diagram of the electromagnetic radiation measured at oscilloscope and 30 cm away from the high voltage generator. This EMR is mainly present in the frequency range below 10 MHz, with a maximum value near 7 MHz. It is therefore a frequency range close to that of radio frequencies emitted by cell phones and for which the long-term health risks are still the subject of epidemiological studies [55]. Nevertheless, we are dealing here with relatively short exposure times (typically a few minutes) while having the possibility of moving the high-voltage generator 2 meters away from the patient in order to naturally reduce the EMR or, better, to confine it in a Faraday cage. The elimination of electromagnetic noise is therefore an important but not limitative issue that will have to be addressed to make cold plasma endoscopy a technology that is both reliable and in line with endoscopists' expectations.

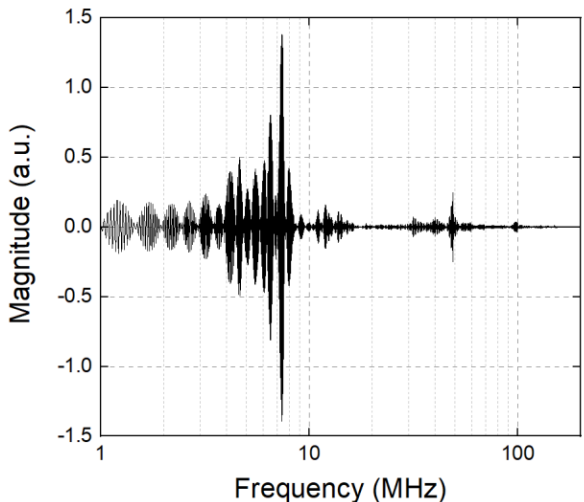

*Figure 12. Spectrum of the electromagnetic radiation emitted by the high voltage power supply and measured 30 cm distance.*

Another expected side effect is the occurrence of eddy currents if the plasma catheter is powered by nanopulsed excitation. As shown in Figure 11c, the current circulating in the auxiliary wire creates a magnetic induction $\vec{B}(r,t) = B(r,t).\vec{e}_\theta$. Its circular field lines can pass through the internal structures of the duodenoscope, especially the closed metallic contours which are both numerous and intersecting. For the sake of clarity, a single closed conductive contour is considered in Figure 11c, made of two parallel angulation wires and the proximal/distal metallic elements integrated into the duodenoscope. This contour is associated with an oriented surface $\vec{S} = \iint dS.\vec{n}$ and the $\vec{B}.\vec{S}$ scalar product corresponds to the magnetic flux $\Phi(t)$ given by the relation {5}. According to the Lenz-Faraday law, a temporal variation of $\Phi(t)$ through the contour of resistance R, generates an electromotive force $e_{ind}(t)$ at the terminals of this contour which is created to oppose the variation of $\Phi(t)$. As indicated in equation {6}, $e_{ind}(t)$ is associated with an induced current $i_{ind}(t)$ which is called "eddy current" if it circulates in a closed metallic contour. This type of current can have several undesirable effects, in particular the loss of power and the heating of certain conductors (energy dissipation by Joule effect).

$$\Phi(t) = \iint_\Sigma \vec{B}(t).dS.\vec{n} \quad \{5\}$$

$$i_{ind}(t) = \frac{e_{ind}(t)}{R} = -\frac{1}{R}.\frac{d\Phi}{dt} \quad \{6\}$$

**(c) Safety of the electrical parameters**

The risks associated with electric current in the context of a cold plasma endoscopy procedure must be clearly identified. It is known that death by electric shock results primarily from ventricular fibrillation and secondarily from asphyxia or cardiac arrest. Table 3 recalls the effects of AC current at 50-60 Hz on the human body [56]. It turns out that the risks become significant when values as high as 20 mA are reached for contact times of 1 min, driving to tetanization of the rib cage. Consequently, and at first sight, the value of 1 Ampere obtained in Figure 10e might suggest that cold plasma endoscopy is not suitable for humans. This conclusion is erroneous, first because the effects reported in Table 3 concern only alternating currents of 50-60 Hz, and second because the biological effects of electricity do not depend solely on the frequency of the current: the type of current (AC, DC, etc.) and the wave profile are of major importance.

For AC currents at frequency higher than 50-60 Hz, the ventricular fibrillation threshold moves away as the frequency of the current increases. This phenomenon can be understood by defining a frequency factor ($\psi$) as in equation {7} and by plotting it versus frequency, as shown in Figure 13a following standard IEC TS 60479-1.

$$\psi = \frac{Fibrillation\ threshold\ at\ f_X(Hz)}{Fibrillation\ threshold\ at\ 50\ (or\ 60)Hz} \quad \{7\}$$

$$w_{pulse} = \frac{D_{Cycle}}{f_{rep}} \quad \{8\}$$

For example, it is known that the physiological effects of a 20 mA current at 50 Hz correspond to a tetanization of the rib cage. If one seeks to obtain the same physiological effect at 1000 Hz, then the amplitude of the current will have to be greater. To find out this value, refer to Figure 13a, which indicates that $\psi$ = 14 for $f_X$ = 1000 Hz. In other words, the same physiological effect will be triggered by a current of frequency 1000 Hz as if its amplitude is 20 × 14 = 280 mA. Similarly, if a sinusoidal current at 10 kHz is delivered, then this threshold increases to 2.8 A. The human body can therefore withstand higher current values as its frequency increases. Within the framework of our experiments, the voltage (and thus the current) is delivered at 10 kHz although its profile is not sinusoidal but pulsed. This means that the current profile must be considered in addition to the frequency, especially the width of the current peaks. The pulse width ($w_{pulse}$) is defined by equation {8} and has a value of 1% / 10 kHz = 1 µs, as graphically confirmed in Figure 10e. To know if it is dangerous to expose a human body to (1 A; 1 µs) current pulses, it is necessary to comply with IEC standards on medical electrical equipment [56]. For short pulses, the standard indicates the energy as the main factor to determine if fibrillation may occur. Therefore, for pulses of various profiles (rectangular, half sinusoidal cycle, decaying exponential) the key parameters are dc peak value (rectangular pulse), rms and peak value (sinus pulse), rms calculated over 3τ and peak value (exponential pulse).

According to IEC standards, fibrillation risks can be classified into four categories: zero, low, medium and high. They can be represented by 4 zones in a graph representing the duration of current pulses (between 0.1 and 100 A) as a function of their amplitude (between 0.1 and 10 ms), as proposed in Figure 13b. The pulsed current measured on the PPBT model in Figure 10e can be reported in Figure 13b as a single point whose coordinates are (1 A; 1 µs). Even if this point falls outside of these 4 zones, the safety of the measured current can be unambiguously assessed. Indeed, the Figure 13b clearly shows that pulses of 1 A and 1 ms width are completely innocuous. Consequently, pulses of 1 µs, thus 1000 times shorter, will be totally negligible. This means that the transferred plasma catheter can be used safely under the experimental conditions tested.

Finally, it is also worth mentioning that our teams (LPP & CRSA) have previously conducted preclinical studies using atmospheric pressure plasma jets that were non-endoscopic devices to treat ectopically-inoculated cholangiocarcinoma tumors in mouse models [31], [32]. At that time, we used the same electrical parameters described in the current study (8-10 kV amplitude, 5 kHz frequency and 1% duty cycle) in order to induce antitumor responses without deleterious effects on the exposed tissues. Our treatment times were typically a few minutes, which is consistent with the recommendations supported by [57] to prevent severe damages.

| Current amplitude | Contact time | Effects |
|---|---|---|
| 0.5 - 1.0 mA | - | Perception threshold according to skin state |
| 8 mA | - | Shock when touching, sudden reactions |
| 10 mA | 4 min | Contraction of limb muscles & Long-lasting muscle twitches |
| 20 mA | 60 s | Beginning of tetanization of the rib cage |
| 30 mA | 30 s | Ventilatory paralysis |
| 40 mA | 3 s | Ventricular fibrillation |
| 75 mA | 1 s | Ventricular fibrillation |
| 300 mA | 110 ms | Ventilatory paralysis |
| 500 mA | 100 ms | Ventricular fibrillation |
| 1000 mA | 25 ms | Cardiac arrest |

*Table 3. Physiological effects of AC current at 50-60 Hz on the human body (for different current magnitudes and contact times).*

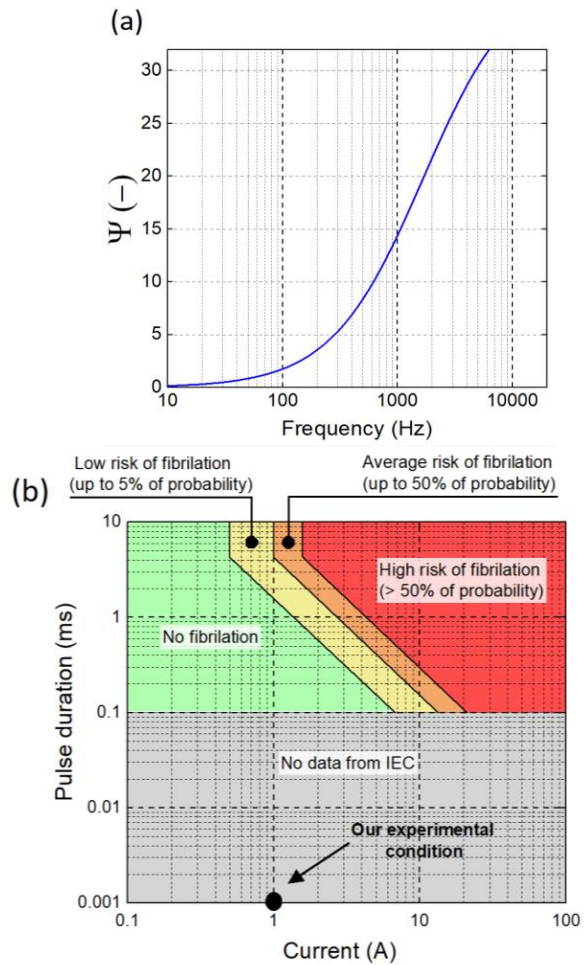

*Figure 13. Thresholds of ventricular fibrillation determined through two graphs: (a) Frequency factor versus frequency of a periodic current applied to human body (b) Pulse duration of a periodic current as a function of its amplitude to highlight fibrillation risks.*

### III.3.3. Thermal properties

As the electrical power deposited in the PPBT model is a few hundred mW (and not a few mW as in the AHBT model), one may wonder whether the conversion of electrical power into calorific power increases too excessively the temperature of the internal tissues. To verify this hypothesis, we focus on the region of the PPBT model comprising the common bile duct and the choledoc (Figure 14a). Thermographic imaging is performed in real time on this region with the catheter inserted into the bile duct and in which helium flows without being ionized. The infrared image in Figure 14b shows that the overall temperature of the biliary tree is around 16.5 °C. This value is slightly lower than the ambient temperature (19 °C) because the PPBT model has been taken out of the freezer (– 20°C) only 12 hours before the experiment. Metal alligator clips are at slightly higher temperatures because they are in thermal equilibrium with the environment. The 14°C zone corresponds to a local cooling of the non-ionized helium flow in the internal tissues.

The variation of the external temperature of the bile ducts can be monitored before, during and after exposure to cold plasma, as seen in Figure 14c. This temperature is measured at a specific point of the infrared map and which corresponds to the most penalizing heating conditions. It turns out that:

(i) Before plasma exposure, the tissue temperature is 16.5°C.
(ii) During a plasma exposure time of 10 min, the temperature increases by 4°C and is therefore completely bearable by the human body. This increase is non-linear: it starts sharply in the first few minutes and continues asymptotically. Since this temperature is measured outside (and not inside) the bile ducts, it may be considered as underestimated. However, such thermal discrepancies are negligible since the thickness of a bile duct is only 1.5 mm.
(iii) After plasma exposure (t ≥ 13.3 min), an immediate and significant drop of tissue temperature is measured. Hence, a decay of 3 °C is reached in less than 3 minutes, so that the temperature is then stabilized at a value close to 17°C.

To ensure that cold plasma exposure does not induce deleterious effects in the bile ducts, the appearance of their inner walls can be compared by cutting them longitudinally with a scalpel. Two cases are considered: PPBT model not exposed to plasma (Figure 14d) and PPBT model after 20 minutes of plasma exposure (Figure 14e). In both cases and in a very qualitative way, the tissues retain the same shiny appearance. In addition, there is no dry, ablated, reddened or blackened area: no damage is visible to the naked eye.

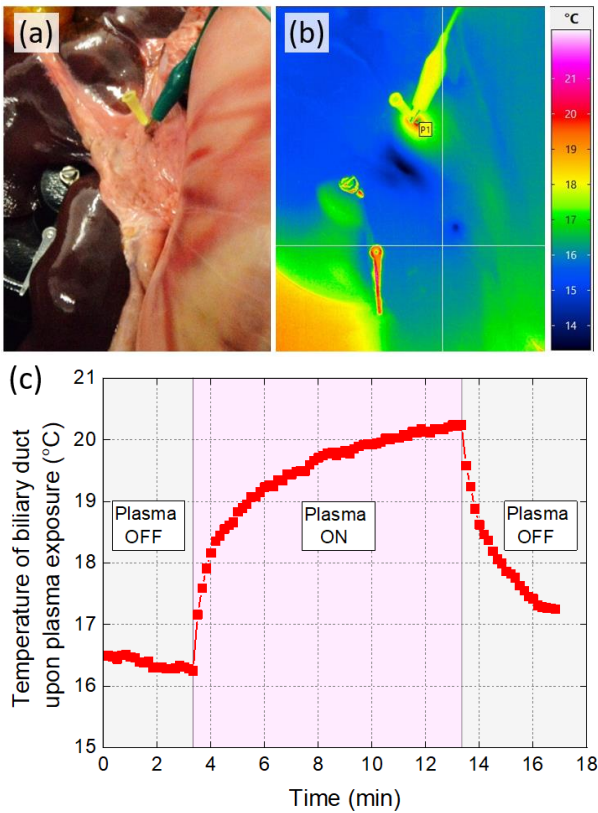

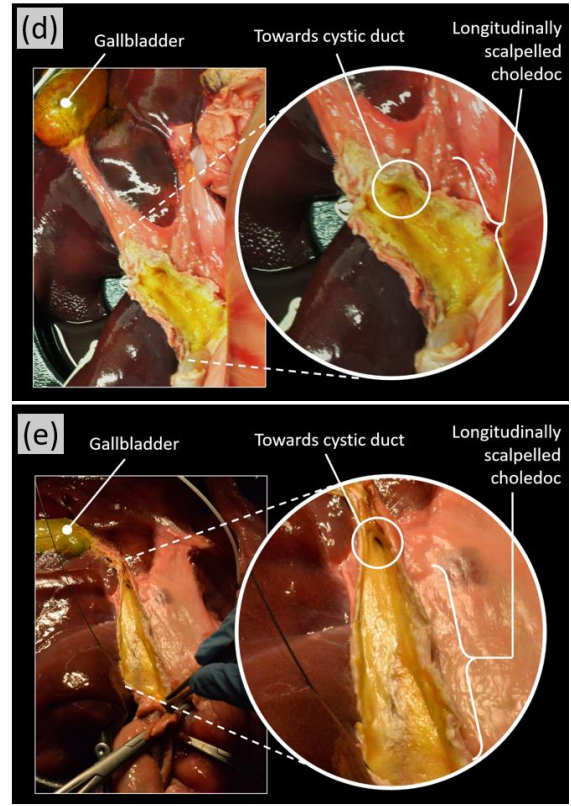

*Figure 14. (a) Photograph of part of the porcine biliary tree (b) infrared picture of (a) showing the absence of overheated regions after 1 min of plasma exposure, (c) Bile duct outer wall temperature as a function of time. Monitoring is achieved before, during and after plasma exposure, (d) Photograph and magnified view of healthy bile ducts from a porcine model, scalped longitudinally to reveal the surface state of the internal walls without exposure to plasma, (e) after plasma exposure (20 minutes).*

## III.4. Therapeutic potential on cholangiocarcinoma (*in vitro* studies)

Finally, we performed *in vitro* studies on human CCA cells to analyze the antitumor effects of the plasma catheter. For this, we evaluated the effects of CAP exposure on the viability of two human CCA cell lines, EGI-1 and HuCCT-1. Briefly, cells were directly treated with using the plasma catheter during 0.5, 1, 2, 3 and 5 min and cell viability was assessed 72 hours post-treatment. We notice that such a treatment induces a significant decrease in the viability of CCA cells and this effect becomes stronger for longer plasma exposure times for both cell lines (Figure 15a). As an example, Figure 15b shows photographs of the HuCCT-1 tumor cells taken 72 hours after several plasma exposure times. Altogether, these results demonstrate that the cold plasma catheter can induce antitumor effects on *in vitro* experimental models of human CCA. This result is consistent with the work that we have already published on CCA treated with non-endoscopic plasma sources, whether *in vitro* or *in vivo* models [31], [32]. Although the question of plasma-triggered biological selectivity (i.e. the ability to kill tumor cells while preserving non-tumor cells) is not addressed in this work, we have already demonstrated this effect by exposing CCA tumor lines and healthy patient-derived liver cells (hepatocytes) to an atmospheric pressure plasma jet [31]. Similar effects can be expected using the transferred plasma catheter.

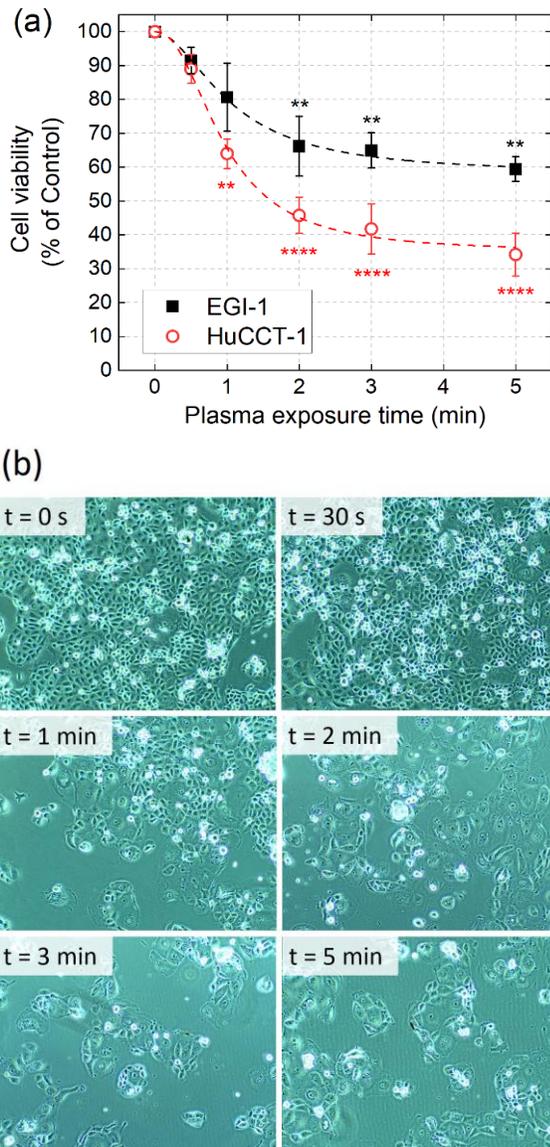

*Figure 15. (a) Effects of CAP on the viability of EGI-1 and HuCCT-1 CCA cells. CCA cells were exposed to CAP for 0.5, 1, 2, 3 and 5 min and cell viability was measured 72 h post-treatment using the crystal violet assay. Values are expressed as means ± SEM from at least 3 cultures. \*, p < 0.05; \*\*, p < 0.01; \*\*\*, p < 0.001; \*\*\*\*, p < 0.0001; as compared with untreated CCA cells. SEM, standard error of the mean, (b) Photographs of the HuCCT-1 tumor cells taken 72h after plasma exposure (Magnification X10).*

## IV. CONCLUSION

After evaluating the performances and limitations of the therapeutic options currently used to treat cholangiocarcinoma, we have shown the existence of a technological niche for the "cold plasma" approach, provided the innovation of *ad hoc* plasma sources. Among the different catheter configurations that can be endoscopically applied to humans, the transferred plasma catheter has been selected owing to its simple design and ease of use. The particularity of this device is that cold plasma is generated in a main chamber (located outside the subject to be treated) and then transferred along a large flexible capillary containing an auxiliary wire. Thus, the plasma can propagate to the distal end of the catheter to form a plasma plume of a few mm-cm. We have demonstrated that this auxiliary wire behaves as an 80-kΩ impedance transmission line and not as a floating electrode because its length ($L_{wire}$ = 2 m) is not negligible compared with the spatial period of the voltage signal (λ = 10 m).

Since the transferred plasma catheter cannot be directly applied to a patient, several intermediate steps have been completed in this work. In a first experimental campaign, the device has been applied to an artificial human biliary tree model (AHBT) supplemented by a tumor insert with the same topographical and electrical characteristics as a solid cholangiocarcinoma tumor. While the cold plasma is confined in an artificial biliary tree, we have measured voltage and current values as low as 3.98 $V_{RMS}$ and 1.19 $mA_{RMS}$ respectively. Besides, the tissues directly exposed to the plasma presented a temperature never exceeding 28 °C. This first experimental campaign has demonstrated that cold plasma can be safely applied on the AHBT model, without electrical or thermal risks.

A second experimental campaign has been carried out at the Paris School of Surgery using conventional ERCP conditions to get as close as possible to a clinical case. The cold plasma catheter has been placed in a conventional duodenoscope which was itself inserted in a new model: a post-mortem porcine biliary tree model (PPBT). Plasma-ERCP has been performed by Dr. Marine Camus, professional endoscopist at Saint-Antoine Hospital. This feasibility study has shown that the internal tissues directly exposed to the plasma do not suffer any thermal damage, the maximum temperature increase being of only 4°C before/during plasma exposure. The electrical analysis shows current and voltage values of about 1 A and 200 V respectively. If these values seem too high for a sage application onto humans, one must however underline that these signals correspond to very short pulses (widths of 1 μs), which means that they do not induce the same physiological effects on the human body as usual AC currents at 50-60 Hz. Here, the IEC standards show that the current values measured on the PPBT model fall into the zero-risk category, (i.e. absence of ventricular fibrillation, cardiac fibrillation, etc.) so that the treatment would be totally painless.

Finally, the antitumor effects of the transferred plasma catheter have been demonstrated in an *in vitro* study performed on human CCA cell lines. Tumor cell viability has decreased by approximately 50% for a plasma exposure time of only 5 minutes.

This methodological study has allowed us to set important milestones in the perspective of using cold plasma endoscopy as a future conventional therapeutic option to treat cholangiocarcinoma in patients. Until then, other steps will be achieved, in particular experimental campaigns on alive porcine models followed by a clinical study. In parallel, questions dealing with biological selectivity of cold plasma (i.e. killing the tumor cells while preserving the non-tumor ones) will be investigated.

# V. ACKNOWLEDGEMENTS

The authors would like to thank Josette Legagneux and her team for managing the logistics of our experimental campaigns at the Ecole de Chirurgie de Paris, as well as Marion Chartier and Gregoire Salin for participating in these campaigns. This work received financial state aid as part of the ASCLEPIOS project funded by the Cancéropôle Ile-De-France (Ref. 193602), the PROMISE project funded by SIRIC CURAMUS (ref. 195741); the $^{PF2}$ABIOMEDE platform co-funded by « Région Ile-de-France » (Sesame, Ref. 16016309) and Sorbonne Université (technological platforms funding). LF is supported by ITMO Cancer of Aviesan on funds administered by Inserm (2021). AP, MC and LF are part of a research team supported by the Fondation pour la Recherche Médicale (Equipe FRM 2020 n°EQU202003010517). LF and AP are members of the European Network for the Study of Cholangiocarcinoma (ENS-CCA) and of the European H2020 COST Action CA18122. This article/publication is based upon work from COST Action CA20114 PlasTHER "Therapeutical Applications of Cold Plasmas", supported by COST.

# VI. DATA ACCESS STATEMENT

The data that support the findings of this study are available upon reasonable request from the authors.

# VII. CONFLICT OF INTEREST DECLARATION

Conflict of Interest declaration: The authors declare that they have NO affiliations with or involvement in any organization or entity with any financial interest in the subject matter or materials discussed in this manuscript.